\documentclass[prl,twocolumn,superscriptaddress,showpacs,floatfix]{revtex4-2}

\usepackage[utf8]{inputenc}
\usepackage{graphicx}
\usepackage{dcolumn}
\usepackage{braket}
\usepackage{bm}
\usepackage{amsfonts} 
\usepackage{amsmath} 
\usepackage[bookmarksnumbered,pdfpagelabels=true,plainpages=false,colorlinks=true,linkcolor=blue,citecolor=blue,urlcolor=blue]{hyperref}
\usepackage[bookmarksnumbered,pdfpagelabels=true,plainpages=false,colorlinks=true,linkcolor=blue,citecolor=blue,urlcolor=blue]{hyperref}

\def \usach {Departamento de F\'isica, Universidad de Santiago de Chile, 9170124, Santiago, Chile.}
\def \cedenna {Centro  de Nanociencia y Nanotecnología CEDENNA, Avda. Ecuador 3493, Santiago, Chile.}
\def \fcfm {Departamento de F{\'i}sica, Facultad de Ciencias Físicas y Matemáticas, Universidad de Chile, Santiago, Chile.}

\def \utarapaca {Departamento de Física, Facultad de Ciencias, Universidad de Tarapacá, Casilla 7-D, Arica, Chile.}

\begin{document}

 \title{Antiferron Modes in Ferroelectric Materials}

\author{David Galvez-Poblete}
\email{david.galvez.p@usach.cl}
\affiliation{\usach}
\affiliation{\cedenna}

\author{Mario A. Castro}
\affiliation{\fcfm}

\author{Roberto E. Troncoso}
\affiliation{\utarapaca}

\author{Guillermo Romero}
\affiliation{\usach}
\affiliation{\cedenna}

\author{Alvaro S. Nunez}
\affiliation{\fcfm}

\author{Sebastian Allende}
\affiliation{\usach}
\affiliation{\cedenna}

\begin{abstract}
We introduce the concept of antiferron modes in ferroelectric materials as dynamically stabilized collective excitations over inverted polarization states that decrease the system energy. While ferrons represent quantized oscillations around the stable polarization minimum, antiferrons require dynamic stabilization via high-frequency driving. Using a generalized Landau-Ginzburg-Devonshire framework, we derive the effective curvature corrections from external driving, demonstrate the conditions for stabilizing metastable wells, and present the quantized Hamiltonian. Antiferrons could be a promising candidate for developing electrical sensing devices, offering tunable, dynamically controllable excitations with high sensitivity to external electric fields.
\end{abstract}

\maketitle

\section*{Introduction}

Ferroelectric materials inherently possess broken inversion symmetry and exhibit anharmonic potential energy landscapes that support rich collective dynamics of the polarization field \cite{Scott2007,Resta,Martin2016,Setter2006}. Within the phenomenological Landau-Ginzburg-Devonshire (LGD) framework \cite{Devonshire1949,Devonshire1951,Chandra,Levanyuk2020}, the free energy features a characteristic double-well potential arising from nonlinear terms in the polarization. Small oscillations around the stable minima of this potential correspond to quantized collective modes with electric dipole moments; these are known as \emph{ferrons} \cite{Tang2022,Tang2024,Zhou2023}, serving as the ferroelectric analogs of magnons in magnetically ordered systems \cite{Troncoso2011,Kruglyak2010,Chen2025,Bauer2023,Chumak2015}.

Analogous to the concept of \emph{antimagnons}~\cite{Harms2024}, which represents excitations around metastable inverted spin configurations stabilized by spin-transfer torques or external fields, we propose the existence of \emph{antiferron} modes in ferroelectrics. Antiferrons are envisioned as coherent collective excitations occurring over a metastable or intrinsically unstable inverted polarization state, and they decrease the energy of the system. Such a state is not naturally stable but can be dynamically stabilized through externally applied high-frequency driving fields or injected polarization currents, effectively reshaping the potential landscape via a Kapitza-like mechanism~\cite{kapitza1951pendulum, Butikov2001}.

This dynamic stabilization enables \emph{external control} over the creation and annihilation of antiferron modes on demand, allowing the polarization landscape to be reconfigured in situ. Unlike ferrons, which exist permanently in the ferroelectric phase with fixed properties, antiferrons can be switched on or off, providing programmable control over their presence, frequency, and spatial localization. This offers new opportunities for reconfigurable ferroelectric devices~\cite{Ram2023,Wu2023,Xu2022}, dynamic modulation of dielectric properties~\cite{OrdoezPimentel2023,Hoerman2002}, tunable THz emission sources\cite{Wang2022,Subedi2025}, and multistable memory~\cite{Jiang2020,Zhang2024,Wang2024,Feng2025}.

In this work, we introduce and analyze antiferron modes as a novel class of dynamically stabilized collective excitations in ferroelectric materials. We develop a theoretical framework to describe their stabilization via high-frequency fields and polarization currents, characterize their collective dynamics, and discuss potential device concepts leveraging their unique combination of dynamic programmability, spectral tunability, and spatial control.

\section*{Generalized LGD Framework with Source}

To describe antiferron modes as collective excitations, we adopt the generalized Landau-Ginzburg-Devonshire (LGD) framework in one spatial dimension, including spatial variations of the polarization field \( P(x,t) \)\cite{Cao2008, Bain2017}. The free energy functional is:
\begin{equation}
F[P(x)] = \int dx \, \mathcal{F}(P, \partial_x P, t),
\end{equation}
where the free energy density \(\mathcal{F}\) is given by:
\begin{equation}
\mathcal{F} = \frac{a}{2} P^2 + \frac{b}{4} P^4 + \frac{c}{6}P^6 + \frac{D}{2}\left(\frac{\partial P}{\partial x}\right)^2 - E(t) P.
\end{equation}

Here, the gradient term \(\frac{D}{2}(\partial_x P)^2\) penalizes sharp spatial variations, representing domain-wall energy and enabling spatially coherent collective modes. We introduce the Lagrangian density for the polarization field:
\begin{equation}
\mathcal{L}(P, \dot{P}, t) = \frac{\rho}{2} \dot{P}^2 - \mathcal{F}(P, \partial_x P, t),
\end{equation}

where \(\rho\) is the effective inertia parameter per unit length. The equation of motion for the system is given by the Landau-Khalatnikov-Tani equation \cite{Ishibashi1989,Widom2010}:
\begin{equation}
    \rho \frac{\partial^2}{\partial t^2}P = - \frac{\delta \mathcal{F}}{\delta P}+ J_p,
\end{equation}

where $J_p$ represents an externally applied current or source term acting on the system. 

To determine the stable points, and thus the equilibrium configuration of the system, we analyze the first derivative of the free energy functional:
\begin{equation}
    \frac{\partial \mathcal{F}}{\partial P}\big|_{P_e} =0 \to aP + bP^3+cP^5 -E(t)  =0,
\end{equation}
where we have considered only uniform configurations as solutions for simplicity, and no external currents have been included. In the case $E(t)=0$, the system exhibits two stable configurations at $+P_o$ and $-P_o$, and an unstable point at $P_u =0 $. The previous results require the quadratic coefficient to satisfy $a<0$.

If an external electric field $E(t) \neq 0$ is applied, the symmetry of the system is broken, resulting in a single global minimum and, consequently, a unique stable configuration. Moreover, depending on the magnitude of the applied electric field, the system may exhibit a local minimum and an unstable point, or only an inflection point. 

Linear excitations, referred to as Ferrons, can propagate around the stable configuration. These excitations follow a dispersion relation in the real domain and correspond to energy-increasing fluctuations of the system.

\bigskip
Antiferrons, defined as linear excitations around unstable or metastable configurations that decrease the system's energy, can be observed by analyzing fluctuations about such non-stationary states. To achieve this, the initial requirement is the dynamic stabilization of an intrinsically unstable configuration.
\section*{Dynamical Stabilization of an Unstable configuration}

Let us consider the case $E(t)=0$. In this situation, the unstable point is located at $P_u = 0 $. If we expand the functional energy around this point:
\begin{equation*}
    \mathcal{F}(P_u + \delta P ) = \mathcal{F}(P_u) + \frac{\partial \mathcal{F}}{\partial P}\big|_{P_u}\delta P+ \dfrac{1}{2}\frac{\partial^2 \mathcal{F}}{\partial P^2}\big|_{P_u}\delta P^2 + \mathcal{O}(\delta P^3)
\end{equation*}
Since $P_u$ is an unstable point, the first derivative of $\mathcal{F}$ at this point is zero, and the second derivative is the parameter $a$:
\begin{equation}
    \mathcal{F} (P_u + \delta P )  = \mathcal{F}(P_u)+ \dfrac{a}{2} \delta P^2 + \frac{b}{4} \delta P^4,
\end{equation}
where contributions from higher-order terms have been neglected since $\delta P$ is a small perturbation. 
To dynamically stabilize this point, inspired by the Kapitza effect \cite{kapitza1951pendulum}, we applied a high-frequency effective current:
\begin{equation*}
    J_p = \varepsilon \cos(\Omega t) \delta P
\end{equation*}
With $\Omega \gg \omega_o$, with $\omega_o = \sqrt{|a|/\rho}$. This term can be interpreted as a dynamical modulation of the polarization rigidity, represented by the parameter $a$, which may originate from multiferroic coupling \cite{TongueMagne2021}, optically induced effects \cite{Chen2022}, or strain-mediated mechanisms \cite{Charnaya2014,Salje1990,Damjanovic1998}. With this inclusion, the equation of motion:
\begin{equation}
    \rho \frac{\partial^2}{\partial t^2} \delta P = - a \delta P + \varepsilon \cos(\Omega t) \delta P 
\end{equation}
As previously stated, $\delta P$ is assumed to be small, and thus only linear terms are retained in the preceding equation. Since the applied current oscillates at a frequency much higher than the system's natural frequency scale, we introduce a perturbative approach based on the separation of fast and slow time scales:
\begin{equation*}
    \delta P(t) = \delta P_{slow} (t)+ \delta P_{fast} (t)
\end{equation*}
\begin{equation*}
    \delta P_{fast}(t) = \Gamma(t) \cos(\Omega t)
\end{equation*}
Introducing this in the motion's equation:
\begin{multline}
    \rho \ddot{\delta P}_{slow} + \cos(\Omega t) \left[ \rho \ddot{\Gamma}- \rho \Gamma \Omega^2 - \varepsilon\delta P_{slow}+ a\Gamma \right] \\
    - 2 \dot{\Gamma} \Omega \rho \sin(\Omega t) + a \delta P_{slow} - \varepsilon\Gamma \cos^2(\Omega t) =0
\end{multline}
The previous equation can be analyzed by decomposing it into trigonometric components. From the sine and cosine coefficients, we can deduce:
\begin{equation}
 \Gamma = \frac{\varepsilon\delta P_{slow}}{a-\rho \Omega^2} = \xi \delta P_{slow}
\end{equation}
Finally, since $\cos^2(\Omega t)$ oscillates at high frequency, we can take its time average and incorporate its contribution as an effective constant parameter in the equation of motion. Considering $\langle \cos^2(\Omega t)\rangle = 1/2$, for the constant terms we obtain:
\begin{equation}
    \rho \ddot{\delta P}_{slow} = -a \delta P_{slow} + \frac{\varepsilon\Gamma}{2} = - a_{\text{eff}} \delta P_{slow}
\end{equation}
\begin{equation*}
    a_{\text{eff}} = a - \frac{\varepsilon^2}{2 (a - \rho \Omega^2)}
\end{equation*}
In summary, the Kapitza-like mechanism introduced here acts analogously to a Floquet procedure for periodically driven systems\cite{Goldman2014,Bukov2015,Eckardt2015}. The high-frequency modulation generates an effective Floquet Hamiltonian in which $a_{\text{eff}}$ governs the slow dynamics. To achieve metastability in this configuration, we require that $a_{\text{eff}} > 0$. Since $a<0 \to a = - |a|$, the metastability condition:
\begin{equation}
\frac{\varepsilon^2}{2(|a|+ \rho \Omega^2)} - |a| >0    
\end{equation}
Furthermore, the effect of this dynamics on the quartic coefficient $b$ can be evaluated. Inserting this expression for $\delta P$ into the quartic contribution to the free energy yields:
\begin{equation}
    b_{\text{eff}} = b \left(1+ 4 \xi cos(\Omega t ) + 6 \xi^2 cos^2(\Omega t) + \mathcal{O}(\xi^3) \right)
\end{equation}

Since $\xi<<1$, higher-order terms in $\xi$ can be neglected. By averaging the oscillatory components, we obtain:
\begin{equation}
    b_{\text{eff}} = b (1+3 \xi^2)
\end{equation}

\begin{figure}
    \centering \includegraphics[width=1.0\linewidth]{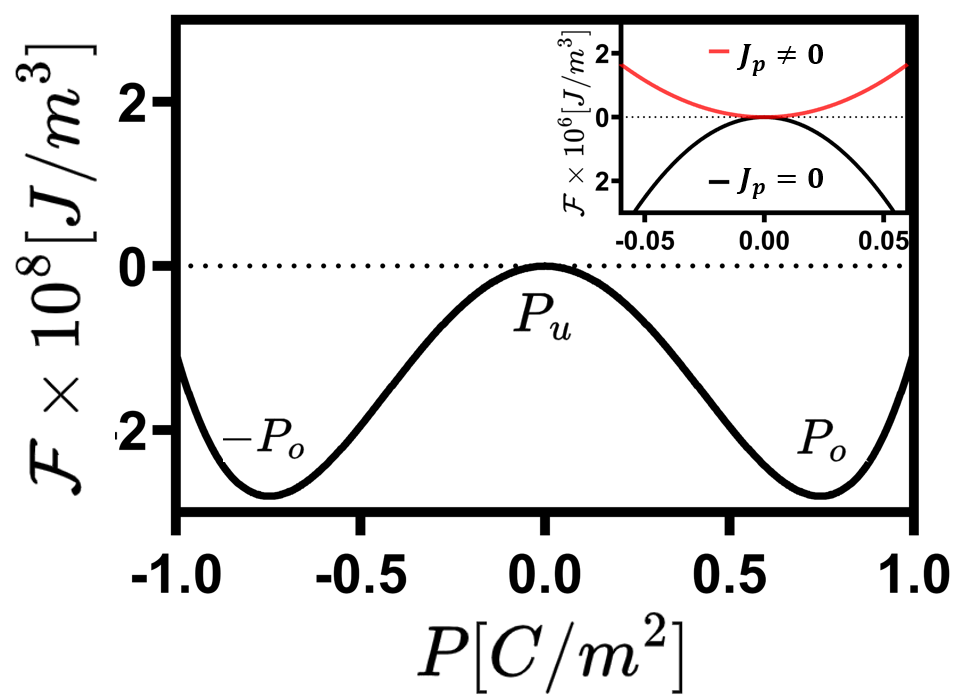}
    \caption{Free energy of ferroelectric system. $P_o$ and $-P_o$ corresponds to the stable minima of the system and $P_u$ denotes the unstable configuration at the local maximum. The inset show the dynamical stabilization of the unstable configuration with $J_p = \varepsilon cos(\Omega t) \delta P$, with $\varepsilon = 3.45\times10^{10} Jm/C^2$ and $\Omega = 10 \omega_o$.}
    \label{fig1}
\end{figure}

The calculations were performed using the ferroelectric parameters of $\text{LiNbO}_3$\cite{Scrymgeour2005}, $a = -2.012\times 10^9 J m/C^2$, $b = 3.608\times 10^9 J m^5/C^4$, $c=0$, $D = 5.39\times 10^{-10}Jm^3/C^2$ and $\rho = 1.81\times 10^{-18} J m s^2/C^2$. However, the analysis is general and can be extended to other ferroelectric materials. In Fig.\ref{fig1}, we show the free energy density as a function of the polarization $P$. As previously discussed, for $J_p =0$, the system exhibits two symmetric minima at $+P_o$ and $-P_o$, along with an unstable configuration at $P_u$. However, when a nonzero current term $J_p\neq 0$ is introduced, with intensity and frequency consistent with the stabilization condition derived earlier, the unstable configuration becomes dynamically stabilized. This stabilization is reflected in the change of curvature, as shown in the inset of Fig.\ref{fig1}.

\section*{Linear anti-excitations: Antiferrons}

From the previous section, we obtained the effective equation of motion governing the slow component of the perturbation. We now introduce linear excitations as this perturbative contribution. Since these excitations may exhibit spatial dependence we reintroduce the gradient term into the equation of motion:
\begin{equation}
       \rho \delta\ddot{ P}_{slow}  = - a_{\text{eff}} \delta P_{slow}  + D \nabla^2 (\delta P_{slow})
\end{equation}
Assuming a linear excitation of the form $\delta P_{slow} = A e^{i(kx-\omega t)}$ \cite{Kittel2004-hd} and inserting it into the effective equation of motion, we derive the corresponding dispersion relation:
\begin{equation}
    \omega^2 = \frac{1}{\rho} \left( a_{\text{eff}} + D k^2 \right)
\end{equation}
Since $a_{\text{eff}}> 0$, the resulting spectrum describes linear and non-divergent excitations, analogous to ferrons in the globally stable configuration. However, if we evaluate the energy contribution of these excitations to the system:
\begin{equation}
 E(\delta P_{slow}) = \frac{1}{2}|A|^2 \left( \rho\omega^2 + a + Dk^2\right)
\end{equation}
Replacing the excitations frequency obtained:
\begin{equation}
    E(\delta P_{slow}) =  \frac{1}{2}|A|^2 \left( 2 a + 2Dk^2  - \frac{\varepsilon^2}{2(a-\rho \Omega^2)}\right)
\end{equation}
From this last expression, the condition to have negative energy excitations:
\begin{equation}
    \frac{\varepsilon^2}{4(|a|+ \rho \Omega^2) }+Dk^2 - |a| <0
\end{equation}
Therefore, under the previous condition and assuming metastability, the system supports linear excitations with negative energy around the metastable configuration, identified as antiferrons. Moreover, this condition allows us to define a critical wavenumber $k_c$:
\begin{equation}
    k_c = \left(\frac{|a|}{D}- \frac{\varepsilon^2}{4D(|a| + \rho \Omega^2)} \right)^{1/2}
\end{equation}

Such that for $k>k_c$, the collective excitations increase the energy of the system, effectively behaving as ferron-like modes around $P_u$. 

\begin{figure}[h!]
    \centering
    \includegraphics[width=1.0\linewidth]{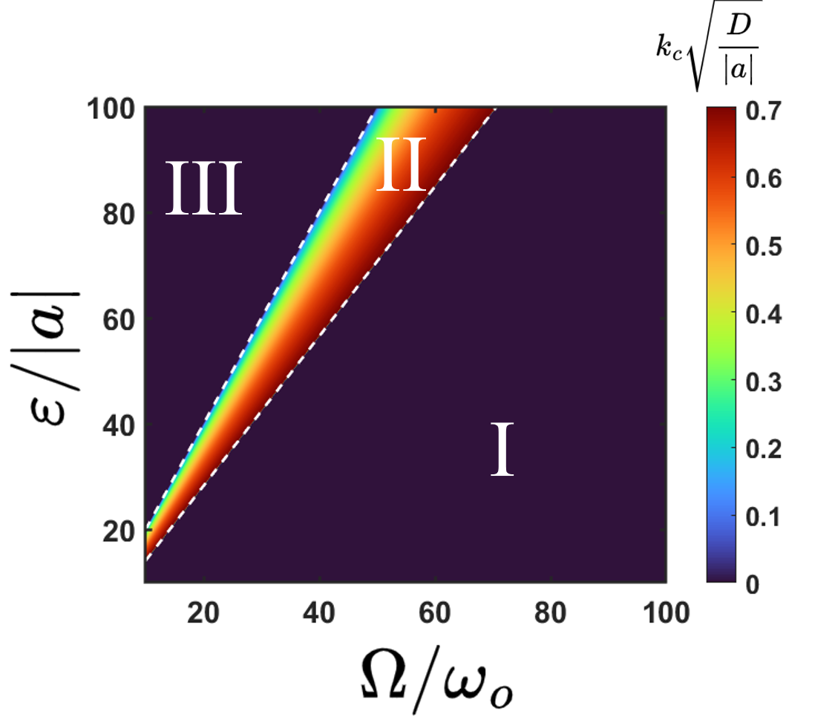}
    \caption{Antiferron modes: stabilization diagram as a function of driving amplitude $\varepsilon$ and frequency $\Omega$. The diagram displays three regions. (I) corresponds to driving parameters insufficient to stabilize the inverted state, with $a_{\text{eff}}<0$. (II) denotes the regime where dynamic stabilization occurs and antiferron modes emerge. (III) represents the range where the driving fully stabilizes the previously inverted state, with $a_{\text{eff}}>0$, but only supports ferronic excitations.}
    \label{fig2}
\end{figure}

Using Eqs. (11) and (18), we performed a numerical analysis to determine the range of driving amplitudes and frequencies for which the system exhibits antiferron modes.
Fig. \ref{fig2} shows three distinct regions. In Region I, the driving field cannot stabilize the inverted point ($a_\text{eff}<0$), and therefore, the system does not exhibit ferron or antiferron modes.
In Region II, the conditions given by Eqs. (11) and (18) are simultaneously satisfied, and collective excitations that decrease the total energy of the system (antiferron modes) appear. Finally, in Region III, the inverted point is dynamically stabilized , but the collective excitations increase the energy of the system, corresponding to ferron modes.

\begin{figure}
    \centering
    \includegraphics[width=1.0\linewidth]{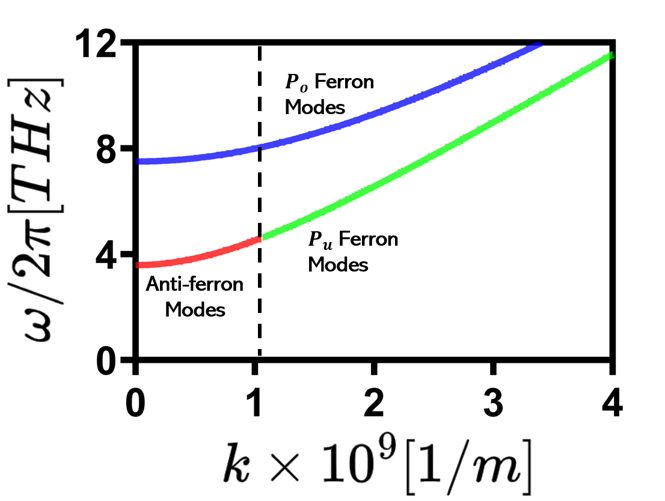}
    \caption{Dispersion relations for collective excitations in ferroelectric system. The red curve denotes antiferron modes, and the green curve represents ferron modes around the unstable point. The blue curve represents the typical ferron modes around the global minima. The dash line represents the $k_c$ that defines if the excitations corresponds to ferron or antiferron modes.}
    \label{fig3}
\end{figure}

In Fig.\ref{fig3}, we observe an energy gap in the low-k regime between the antiferron and ferron modes using the ferroelectric parameters of $\text{LiNbO}_3$\cite{Scrymgeour2005}. This implies that, under the high-frequency driving, it is energetically more favorable for the system to excite antiferrons rather than conventional ferrons.

In practical terms, the driving force $J_P$ may originate from diverse physical mechanisms that modulate the effective curvature of the ferroelectric potential. Possible implementations include multiferroic coupling with a magnetic subsystem, optically induced modulation of the lattice polarization, or strain-mediated coupling through piezoelectric substrates. Additional techniques can be employed to pre-soften the lattice rigidity, for instance, by slightly increasing the temperature below $T_c$. The efficiency of each mechanism depends on the intrinsic stiffness of the ferroelectric material. As an example, in $\text{BaTiO}_3$, a soft ferroelectric, antiferron modes could be realized by a mild pre-heating procedure that reduces the rigidity coefficient. Using the Landau parameter $a(T)= 6.68 \times 10^5 (T-381) Jm/C^2$ \cite{Bell2001},at $T=377K$ one obtains $a_o (377K)= -2.67\times10^6 Jm/C^2$, which corresponds to a natural angular frequency $\omega_o =2\pi \times 0.22$ THz. From equations (11) and (18), we find that a driving term of the form $J_P=15|a|\cos(10\omega_o t)$ is sufficient to induce antiferron modes in this regime. Under a strain-mediated mechanism, the rigidity coefficient is dynamically modified according to $a(t,T=377)=a_o (377K)-2qu(t)$, within a principal-axis approximation, where $q=1.42\times 10^{10}  Jm/C^2$ is the stress-coupling coefficient of $\text{BaTiO}_3$ \cite{Hlinka2006} and $u(t)$  is the time-dependent strain. By comparing this expression with the effective driving amplitude $J_P$, one finds that a dynamic strain with an amplitude of about $0.14\%$ at $\Omega=10 \omega_o = 2\pi\times 2.2$ THz, which is high but experimentally achievable, should be sufficient to observe the antiferron modes. Since this approach may enhance decoherence and dissipation, an alternative strategy is to combine a weaker pre-heating stage with a strain-mediated mechanism and additional pre-infrared pumping \cite{Subedi2014,Frst2011}.

Our one-dimensional model is also fully applicable to a uniaxial ferroelectric derived from a centrosymmetric paraelectric phase, as in the case of $\text{LiNbO}_3$. For the multiaxial case, the analysis can be extended by constructing a free-energy functional that accounts for the different ferroelectric interactions along each principal axis \cite{Scrymgeour2005,Marton2010}. However, deriving the corresponding stability conditions becomes extremely challenging, as the result depends on the specific crystal symmetry and, in general, cannot be obtained analytically. It is expected that, in multiaxial systems, the driving term should have components along several axes, enabling antiferron modes to appear in specific directions. For instance, in a simple cubic system, the driving field must necessarily have three components to achieve dynamic stabilization, and therefore antiferron modes, along the three principal axes. If the driving amplitudes are not identical in all directions and the dynamical stabilization occurs only along certain axes, antiferron modes will propagate exclusively along those stabilized directions, while in the others no propagating modes will exist. In realistic ferroelectric samples, multi-domain configurations are common. The local Landau–Ginzburg–Devonshire description used here remains valid within each domain, so antiferron modes could exist locally.

\section*{Thermal Fluctuations}
All previous analyses and calculations were performed using the Landau free-energy parameters evaluated at room temperature. In order to investigate the influence of thermal effects on the system, we have included dissipation and a Langevin thermal-noise \cite{SIVASUBRAMANIAN2004} term in Eq. (14):
\begin{equation}
    G^{-1}\delta P_{slow} = -3b_{\text{eff}}\delta P_{slow}^3 + E_{th}
\end{equation}
\begin{equation*}
    G^{-1} =  \rho \frac{\partial^2}{\partial t^2} + \gamma\frac{\partial}{\partial t} +a_{\text{eff}}-D\nabla^2
\end{equation*}
Here, $G^{-1}$ denotes the inverse propagator of the linearized part of Eq. (20), while $E_{th}$ is the Langevin thermal term satisfying the fluctuation–dissipation theorem \cite{Landau1996-yt}:
\begin{equation*}
    \langle E_{th} (\textbf{k},\omega) E_{th}(\textbf{k}',\omega ' ) \rangle = \frac{(2\pi)^4 \gamma \hbar \omega \delta(\textbf{k}-\textbf{k'})\delta(\omega-\omega')}{tanh(\hbar \omega /2 k_B T)}
\end{equation*}
In this expression, $E_{th}(\textbf{k},\omega)$ denotes the Fourier transform of the Langevin thermal noise term. From Eq. (20), assuming small thermal fluctuations for $T<T_c$, one obtains:
\begin{equation}
    \delta P_{slow} = -3 b_{\text{eff}} G\delta P_{th}^3 + \delta P_{th}
\end{equation}
With $\delta P_{th} = G E_{th}$ representing the harmonic thermal fluctuations, which satisfy $\langle \delta P_{th} \rangle =0$. In Fourier space, this term can be written as:
\begin{equation*}
    \delta P_{th} (\textbf{k},\omega ) = \frac{E_{th}(\textbf{k},\omega )}{\rho (\omega_\textbf{k}^2-\omega^2 ) -i\gamma \omega }
\end{equation*}
Where $\omega_{\textbf{k}}$ satisfies the expression previously obtained for the antiferron modes. Taking the average over Eq. (21), we find that:
\begin{equation*}
    \langle \delta P_{slow}\rangle \propto \langle \delta P_{th}^3\rangle =0 
\end{equation*}
This result indicates that the thermal fluctuations do not alter the configuration established by the driving at $P_u =0$. Moreover, we can evaluate the influence of these fluctuations on the curvature, and consequently, on the propagation of the antiferron modes. From Eq. (20), we obtain:
\begin{equation}
    \tilde{a} =a_{\text{eff}} + 3 b_{\text{eff}}\langle \delta P_{th}^2\rangle 
\end{equation}
Using the fluctuation-dissipation theorem, and assuming weak dissipation limit ($\gamma \ll \rho \omega_\textbf{k}$), the term $\langle \delta P_{slow}^2 \rangle$ is given by: 
\begin{equation}
    \langle \delta P_{th}^2 \rangle = \frac{\hbar}{4\rho \pi^2} \int dk \frac{k^2}{\omega_k} \text{coth}\left( \frac{\hbar \omega_k}{2k_B T} \right)
\end{equation}
Here, $\omega_{\textbf{k}}$ now depends on $\tilde{a}$. By solving Eqs. (22) and (23) in a self-consistent manner, we obtain a relative deviation of approximately $3\%$ in $a_{\text{eff}}$ due to the effect of thermal fluctuations at room temperature for $\text{LiNbO}_3$
.
\section*{Quantum Model}

We describe small fluctuations of the polarization field around the dynamically stabilized inverted state:
\begin{equation}
\hat{P}(x,t) = -P_u + \delta \hat{P}(x,t).
\end{equation}

The conjugate momentum operator is defined from the Lagrangian density as:
\begin{equation}   
\hat{\Pi}(x,t) = \frac{\partial \mathcal{L}}{\partial \dot{\delta \hat{P}}(x,t)} = \rho \frac{\partial \delta \hat{P}}{\partial t}.
\end{equation}

This leads to the equal-time canonical commutation relation,$
[\delta \hat{P}(x), \hat{\Pi}(x')] = i\hbar \delta(x - x')$ \cite{Altland2010,Fetter1972}.
Expanding the Landau-Ginzburg-Devonshire free energy density to quartic order in \(\delta \hat{P}\), the effective Hamiltonian density becomes:
\begin{equation}
\mathcal{H} = \frac{1}{2\rho} \hat{\Pi}^2 + \frac{1}{2}a_{\text{eff}} \, (\delta \hat{P})^2 + \frac{D}{2} \left(\frac{\partial \delta \hat{P}}{\partial x}\right)^2  + \frac{b_{\text{eff}}}{4}(\delta \hat{P})^4
\end{equation}

Here \(a_{\text{eff}}\) is the effective curvature incorporating static and dynamically induced stabilization.
We impose periodic boundary conditions over a quantization length \(L\). The fluctuation operator is expanded in plane-wave modes:
\begin{equation*}
\delta \hat{P}(x,t) = \frac{1}{\sqrt{L}} \sum_k \hat{q}_k(t) \, e^{ikx},
\end{equation*}
\[
\hat{\Pi}(x,t) = \frac{1}{\sqrt{L}} \sum_k \hat{p}_k(t) \, e^{ikx}.
\]

These mode operators satisfy $
[\hat{q}_k, \hat{p}_{k'}] = i\hbar \delta_{k,k'}.$ Substituting the mode expansions into the total Hamiltonian:
\begin{equation}
\hat{H}_{\text{quad}} = \sum_k \left[\frac{1}{2\rho} \hat{p}_k \hat{p}_{-k} + \frac{1}{2}a_{\text{eff}} \hat{q}_k \hat{q}_{-k} + \frac{D}{2} k^2 \hat{q}_k \hat{q}_{-k}\right].
\end{equation}

We identify that each mode \(k\) behaves as an independent harmonic oscillator with effective frequency $
\omega_k^2 = \frac{1}{\rho} \left( a_{\text{eff}} + D k^2 \right)
$. For each mode \(k\), we define bosonic operators:
\[
\hat{b}_k = \sqrt{\frac{\rho \omega_k}{2\hbar}} \hat{q}_k + i \sqrt{\frac{1}{2\hbar \rho \omega_k}} \hat{p}_k,
\]

\[
\hat{b}_k^\dagger = \sqrt{\frac{\rho \omega_k}{2\hbar}} \hat{q}_{-k} - i \sqrt{\frac{1}{2\hbar \rho \omega_k}} \hat{p}_{-k}.
\]

These satisfy $
[\hat{b}_k, \hat{b}_{k'}^\dagger] = \delta_{k,k'}.$ Expressed in terms of \(\hat{b}_k, \hat{b}_k^\dagger\), the quadratic Hamiltonian diagonalizes to \cite{Tsvelik2003}:
\begin{equation}   
\hat{H}_{\text{quad}} = \sum_k \hbar \omega_k \left( \hat{b}_k^\dagger \hat{b}_k + \frac{1}{2} \right).
\end{equation}

Each mode corresponds to a quantized antiferron excitation with energy \(\hbar \omega_k\). From the Hamiltonian density, we can also extract the anharmonic part of the Hamiltonian, which is associated with the quartic term:
\begin{equation}
    \hat{H}_{\text{anh}} = \frac{\hbar^2 b_{\text{eff}}}{16 \rho^2} \sum_{k_1+k_2+k_3+k_4=0}  \prod_{j=1}^4 \frac{1}{\omega_{k_j}^{1/2}}\left(b_{k_j} + b_{-k_j}^{\dagger} \right)
\end{equation}
These anharmonic contributions could be relevant for qubit-like behavior \cite{Askerzade2023,Krantz2019,Yang2024}, and could thus render this system of particular interest for quantum technologies.

\section*{Application Proposal}
Consider that a small DC electric field $E_{DC}$ is applied to the system. This addition displaces the unstable point to $P_u' = E_{DC}/a$, and consequently modifies the curvature at this point to $\kappa = a + 3bE_{DC}^2/a^2$, which remains negative under small fields. However if we repeat the previous procedure and include a high-frequency driving term, as shown before, this curvature can be effectively shifted to a positive value, dynamically stabilizing the configuration.
\begin{equation}
    \kappa_{\text{eff}} = \kappa - \frac{\varepsilon^2}{2(\kappa - \rho \Omega^2)} > 0 
\end{equation}
Analogously to the previous case, we can derive the condition required to observe antiferron excitations:
\begin{equation}
    \frac{\varepsilon^2}{4(|\kappa|+\rho \Omega^2)}+Dk^2 - |\kappa| < 0
\end{equation}
With dispersion relation:
\begin{equation}
    \omega ^2 = \frac{1}{\rho}(\kappa_{\text{eff}}+Dk^2)
\end{equation}
Here, the effective curvature $\kappa_{\text{eff}}$ depends directly on the applied electric field, thereby modifying the energy gap required for the emergence of these excitations. This phenomenon may enable the detection of small variations in the electric field by measuring the zero mode of antiferronic excitations.

In analogy with the experimental signatures predicted for antimagnons, the presence of antiferron modes could manifest through superradiant-like amplification phenomena \cite{Errani2025}, observable as enhanced transmission in THz or optical pump–probe experiments.

\section*{Conclusions}
In the present work, we study a one-dimensional ferroelectric system within the Landau-Ginzburg-Devonshire(LGD) formalism, employing the Landau-Khalatnikov-Tani equation to describe the polarization dynamics. We analyze both stable and unstable configurations of the system. In particular, by introducing a high-frequency driving field, we demonstrate the possibility of dynamically stabilizing an otherwise unstable configuration. We derive the conditions that the driving parameters must satisfy to achieve such stabilization. Around this dynamically stabilized point, we propose the emergence of a new class of collective excitations, which we term antiferrons. These excitations arise from the unstable configuration and are characterized by a decrease in the system's energy, We further determine the conditions necessary for the existence of antiferrons and discuss their potential application in sensing small variations in external electric fields. Therefore, antiferrons could be a promising candidate for developing electrical sensing devices, offering tunable, dynamically controllable excitations with high sensitivity to external electric fields.

\section*{Acknowledgement}

SA acknowledges funding from DICYT regular 042431AP and Cedenna CIA250002. ASN acknowledges funding from Fondecyt Regular 1230515 and Cedenna CIA250002. R.E.T. thanks funding from Fondecyt Regular 1230747. M.A.C. acknowledges Proyecto ANID Fondecyt Postdoctorado 3240112.  D. Gálvez-Poblete acknowledges ANID-Subdirección de Capital Humano/Doctorado Nacional/2023-21230818.

\end{document}